\documentstyle[preprint,aps]{revtex}
\begin{document}
\author{Helio V. Fagundes \\
\textit{Instituto de F\'{i}sica Te\'{o}rica - IFT/UNESP - S\~{a}o Paulo,
Brazil}}
\title{On Multiple Images in Closed Universes}

\maketitle

\tightenlines
\begin{abstract}
Universes with multiconnected spatial sections predict multiple images of
cosmic sources. A confusing terminology exists in the naming of these images
as real ones vs. ghosts. Here an attempt is made to clarify the situation. 
\end{abstract}

\section{Introduction}

In a cosmological model with closed, nontrivial space topology, a distant
astronomical source is predicted to produce several images, corresponding to
the diffferent paths its radiation can take to reach the observer on Earth.
See Lachi\`{e}ze-Rey \& Luminet \cite{LaLu} for a review.

Theoretically these multiconnected spatial sections are manifolds $M$ of the
form $\tilde{M}/\Gamma ,$ where $\tilde{M}$ is the universal covering space
of $M,$ and $\Gamma $ is a discrete group of isometries acting freely on $%
\tilde{M}$. This action tesselates $\tilde{M}$ into cells which are copies
of a fundamental polyhedron (FP). The FP has an even number of faces, which
are pairwise congruent; $M$ may be represented by the FP, with members of
these pairs identified (more colloquially, they are \ ``glued together'';
more technically, see Massey \cite{Massey}).

In practice one looks at the FP as the physical space where the observer and
the sources exist, and at its copies as the apparent space of repeated
cosmic images. Hence the tendency to consider those images within the FP as
`real,' those in other cells as `ghosts.' But it also appears reasonable to
call real the nearest image of an object, the other ones being the\ ghosts.
It is the purpose of this paper to show that these two criteria for
classifying multiple images are not always consistent with each other.

\section{A two-dimensional example}

One of the possible universe models with Einstein-de Sitter metric

\[
ds^{2}=c^{2}dt^{2}-(t/t_{0})^{4/3}(dx^{2}+dy^{2}+dz^{2})
\]
has as spatial section the orientable manifold E2 in Fig. 17 of \cite{LaLu}.
A Klein bottle surface $K^{2}$ is imbedded in E2, so we may use this
nonorientable surface as our simplified model of cosmic space. It turns out
that the FP (here fundamental \textit{polygon}) is not unique; it depends on
an arbitrary \textit{basepoint }in its definition: The FP of a manifold $M=%
\tilde{M}/\Gamma $ with basepoint $x\in \tilde{M}$ is the set of points $%
\{y\in \tilde{M};$ \textit{distance}$(y,x)\leq $\textit{\ \ distance}$%
(y,\gamma x),\forall \gamma \in \Gamma \}.$

Let $M$ $=K^{2}$ be the Klein bottle obtained from the square $ABCD$ of side
length $=6$ in Fig. 1, with coordinates $(x,y)$ and basepoint $bp0=(0,0).$
This fundamental polygon will be called FP0. Group $\Gamma $ is generated by
the motions

\begin{eqnarray*}
\gamma _{1} &:&(x,y)\mapsto (x+6,-y) \\
\gamma _{2} &:&(x,y)\mapsto (x,y+6) ,\
\end{eqnarray*}
and $\tilde{M}$ is the Euclidean plane. The identified pairs of sides are $%
AD\Leftrightarrow \gamma _{1}(AD)=CB,$ $AB\Leftrightarrow \gamma _{2}(AB)=DC$%
.

Now we choose another basepoint, $bp1=(2,2),$ and find the corresponding FP
as indicated in Fig. 1. The six images of $bp1$ in the figure ($im1-im6)$
are sufficient to determine the irregular hexagon $EFGHIJ$ as FP1. The
points on line $EF$ are equidistant from $bp1$ and $im1$, and thus belong to
the border; and similarly for the other sides. \ The coordinates of the
vertices are $E=(-1/3,-1)$, $F=(13/3,-1)$, $G=(17/3,1)$, $H=(13/3,5)$, $%
I=(-1/3,5),$ and $J=(-5/3,1)$, and the identifications are $%
EJ\Leftrightarrow \gamma _{1}(EJ)=GF$, $EF\Leftrightarrow \gamma _{2}(EF)=IH$%
, $GH\Leftrightarrow \gamma _{2}\gamma _{1}^{-1}(GH)=IJ.$

\section{Ambiguity of the names `real' and `ghost' images}

In Fig. 2, let the observer's position be $bp1$, and two images of a cosmic
source be located at $p=(-2,-1)$ and $q=\gamma _{1}p=(4,1).$ The nearest
image is $q$, and if the observer is using FP1, then by both criteria at the
end of Sec. I $q$ would be the real image and $p$ a ghost. But if she or he
is using FP0, which is much easier to handle mathematically despite the fact
that the observer is \textit{not} at its center, then by first criterion the
real image is $p$ and $q$ is a ghost, with the opposite holding by the
second rule.

Therefore, in order that the terminology `real' vs. `ghost' (or `source' vs.
`ghost') could be used consistently, one would have to always work with an
FP where the basepoint is at the observer's position. A further advantage of
thus having the nearest images labeled as sources would be in the comparison
with astrophysical data, which are usually richer for nearby objects.

However, the calculations are much simpler in the more symmetrical FP like FP0 
above, or in those FP's which mathematicians say have ``maximum injectivity 
radius'' - see Weeks \cite{JW}, for example. This is what was done by the 
author in \cite {QGAII}, where the FP is a regular icosahedron with basepoint 
at its center, far from the observer's position.

I suggest calling the nearest images just that, while the second nearest
would be second image, and so on.

I am grateful to Conselho Nacional de Desenvolvimento Cient\'{i}fico e
Tecnol\'{o}gico (CNPq - Brazil) for partial financial support.


\begin{thebibliography}{9}
\bibitem{LaLu}  M. Lachi\`{e}ze-Rey and J.-P. Luminet, Phys. Rep. \textbf{254%
}, 135 (1995).

\bibitem{Massey}  W. S. Massey, \textit{Algebraic Topology: An Introduction}
(Springer-Verlag, New York, 1967), Appendix A.

\bibitem{JW}  J. R. Weeks, \textit{SnapPea: A Computer Program for Creating
and Studying Hyperbolic Manifols}, available by anonymous ftp from
geom.umn.edu (latest version 2.5.3, 1998).

\bibitem{QGAII}  H. V. Fagundes, Astrophys. J. \textbf{338}, 618 (1989).
\end{thebibliography}
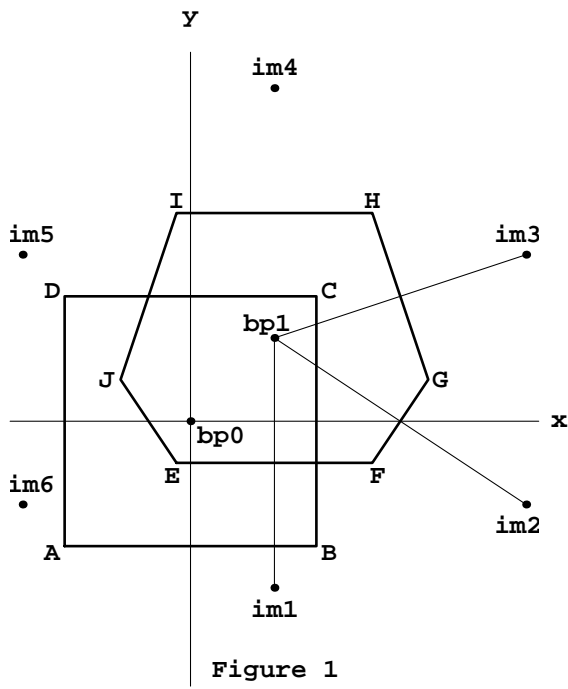
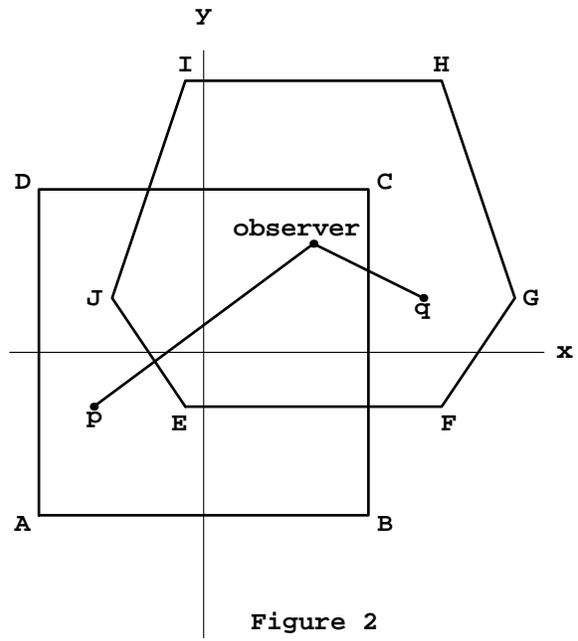
\end{document}